\documentclass[times,twocolumn,review]{elsarticle}

\usepackage{medima}
\usepackage{framed,multirow}

\usepackage{amssymb}
\usepackage{latexsym}

\usepackage{url}
\usepackage{xcolor}

\usepackage{hyperref}

\usepackage{adjustbox}
\usepackage{numprint}
\usepackage{multirow}
\usepackage{booktabs}
\usepackage{svg}
\usepackage{svg-extract}
\usepackage{bm}
\usepackage{amsmath}

\journal{Medical Image Analysis}

\begin{document}

\npdecimalsign{.}
\nprounddigits{3}

\verso{Gonz\'{a}lez et al.}

\begin{frontmatter}

\title{Distance-based detection of out-of-distribution silent failures for Covid-19 lung lesion segmentation}

\author[1]{Camila Gonz\'{a}lez\corref{cor1}}
\cortext[cor1]{Corresponding author: 
  Tel.: +49 6151 155-652;}
\ead{camila.gonzalez@gris.tu-darmstadt.de}
\author[1]{Karol Gotkowski}
\author[1]{Moritz Fuchs}
\author[2]{Andreas Bucher}
\author[2]{Armin Dadras}
\author[2]{Ricarda Fischbach}
\author[2]{Isabel Jasmin Kaltenborn}
\author[1]{Anirban Mukhopadhyay}

\address[1]{Darmstadt University of Technology, Karolinenplatz 5, 64289 Darmstadt, Germany}
\address[2]{Uniklinik Frankfurt, Theodor-Stern-Kai 7, 60590 Frankfurt am Main, Germany}

\begin{abstract} 

Automatic segmentation of ground glass opacities and consolidations in chest computer tomography (CT) scans can potentially ease the burden of radiologists during times of high resource utilisation. However, deep learning models are not trusted in the clinical routine due to \emph{failing silently} on out-of-distribution (OOD) data. We propose a lightweight OOD detection method that leverages the Mahalanobis distance in the feature space and seamlessly integrates into state-of-the-art segmentation pipelines. The simple approach can even augment pre-trained models with clinically relevant uncertainty quantification. We validate our method across four chest CT distribution shifts and two magnetic resonance imaging applications, namely segmentation of the hippocampus and the prostate. Our results show that the proposed method effectively detects far- and near-OOD samples across all explored scenarios.

\end{abstract}


\end{frontmatter}


\section{Introduction}

Automatic segmentation of lung lesions in chest computed tomography (CT) scans could standardise quantification and staging of pulmonary diseases such as Covid-19 and open the way for more effective utilisation of hospital resources. Ground glass opacities (GGOs) and consolidations are characteristic of pulmonary infections onset by the SARS-CoV-2 virus \citep{parekh2020review}. Since the early phases of the pandemic, many institutions have compiled scans from afflicted patients in intensive care, and some initiatives have publicly released cases with ground-truth delineations from expert thorax radiologists \citep{roth2021rapid,ma_jun_2020_3757476,morozov2020mosmeddata}. Deep learning has shown promising results in segmenting these patterns. Particularly the fully-automatic \textit{nnU-Net} \citep{isensee2021nnu} secured top spots \citep{challenge_results} (9 out of 10, including the first) in the leaderboard for the \textit{Covid-19 Lung CT Lesion Segmentation Challenge} \citep{roth2021rapid}.

Unfortunately, models trained with publicly available cohorts may not generalise well to real-world clinical data, thus posing safety issues when deployed without extensive testing and/or quality assurance (QA) protocols. Deep learning models are known to fail for data that diverges from the training distribution \citep{mehrtash2020confidence}; a phenomenon commonly referred to as \emph{domain shift}. This hinders the deployment of AI solutions during the Covid-19 pandemic \citep{hu2020challenges}, as most institutions do not dedicate resources to annotate in-house datasets. There are many potential causes for domain shift, ranging from changes in the acquisition process to naturally shifting patient populations. Some can unknowingly occur within the same institution, rendering even models trained with in-house data unreliable with the passage of time \citep{srivastava2021continual}.

This performance deterioration is visualised in Figure \ref{fig:boxplot_performance} for an nnU-Net trained on data from the \emph{COVID-19 Lung CT Lesion Segmentation Challenge} \citep{roth2021rapid,an2020ct,clark2013cancer}. Featuring 199 cases, 160 of which were used for training, the data pool is much larger than single institutions realistically collect and annotate, considering how time-intensive the process of lung lesion delineation is. The data is also multi-centre and diverse with regard to patient group and acquisition protocol, yet the model fails to generalise to different distribution shifts. Lung lesions do not manifest in large connected components (see Figure \ref{fig:qual}), so it is not trivial for novice radiologists to identify incorrect segmentations.

While we have so far painted a sombre outlook for clinical use of deep learning models, these could still be safely utilised alongside proper quality assurance mechanisms. The problem is that human-performed QA is time-consuming and expensive, ultimately defeating the promise of AI in radiology. On the other hand, automatic methods may be an inexpensive and effective first step in identifying low-quality cases. In particular, reliable \emph{out-of-distribution} (OOD) detection can signal when the model is unsuitable for a patient.

Existing methods for OOD detection or uncertainty quantification either (a) observe the network logits, which often \textit{fail silently} exhibiting plausible behaviour mimicking in-distribution (ID) cases even for novel inputs \citep{hein2019relu} or (b) require special training considerations that reduce their usability, such as a self-supervision loss term or outlier detector. In practice, models are used which exhibit the best performance in the target task. Widely-used segmentation frameworks \emph{are not designed with OOD detection in mind}, and so a method is needed that reliably identifies OOD samples post-training while requiring minimal intervention.

We propose to directly estimate the similarity of new samples to the training distribution in a low-dimensional feature space. A large distance signals that the model has not seen specific activation patterns in the past, and therefore outputs produced from such novel features \emph{cannot be trusted}. Our method \citep{gonzalez2021detecting}, initially presented at MICCAI 2021, is lightweight and requires no changes to the network architecture of the training procedure, allowing it to integrate into complex segmentation pipelines seamlessly. Further, as the distance estimation process follows after training, it can provide clinically-relevant uncertainty scores for pre-trained models. 

Building on our previous work, in the present article we provide more context into our methodology, perform an ablation study on selecting feature maps and considerably extend our evaluation. We validate our proposed method across \emph{four} scenarios with a nnU-Net trained on \emph{Challenge} data.

\begin{enumerate}
    \item For the first setting, we perform inference on the publicly available \emph{Radiopedia} and \emph{Mosmed} datasets. This setting, which we have explored in the past, simulates a \textit{dataset shift} situation where the user does not know exactly which changes are introduced.
    \item Secondly, we apply affine transformations and synthetic artefacts to the ID test data in order to simulate, respectively, geometric changes in the subject population and common quality problems in CT acquisition.
    \item We also evaluate a \emph{diagnostic shift} scenario on an in-house data cohort with 50 Covid-19 and 50 new non-Covid pneumonia patients.
    \item Finally, we carry out a \emph{far-OOD} evaluation where we feed colon and spleen CT examinations from the \emph{Medical Segmentation Decathlon} (MSD) to the model.
\end{enumerate}

In addition, we explore two additional segmentation tasks to assess the transferability of our method to other settings, namely hippocampus and prostate segmentation from, respectively, T1- and T2-weighted Magnetic Resonance Images (MRIs). We also perform experiments on a HighResNet \citep{li2017compactness} architecture, which does not follow the classic encoder-decoder structure.

Our results show that our proposed distance-based method reliably detects out-of-distribution samples that other approaches fail to identify across a wide array of use cases.
\section{Related Work}
Several strategies have shown acceptable OOD detection performance in classification tasks. \emph{Output-based} methods assess the confidence of the logits by estimating their distance from a one-hot encoding. \citet{hendrycks2016baseline} propose using the maximum softmax output as an OOD detection baseline. \citet{guo2017calibration} find that replacing the regular softmax function with a \textit{temperature-scaled} variant produces truer estimates, and \citet{liang2018enhancing} complement this approach by adding perturbations to the network inputs. Similarly, \citet{liu2020energy} use \emph{Energy Scoring} to detect OOD samples in a post-hoc fashion. Given access to explicit OOD samples, training with an energy-based loss can further improve OOD detection. Other methods \citep{hendrycks2019using,lee2018training} instead look at the KL divergence of softmaxed outputs from the uniform distribution.

\emph{Sample-based} Bayesian-inspired techniques \citep{blundell2015weight} consider the divergence between several outputs produced under different conditions as the uncertainty. Commonly-used methods are Monte Carlo Dropout (MC Dropout) \citep{gal2016dropout} and Deep Ensembles \citep{lakshminarayanan2017simple}. The latter usually performs better but requires several models to be trained, whereas MC Dropout can assess uncertainty for any model trained with Dropout layers. \citet{ashukha2019pitfalls} show that Test-Time Augmentation (TTA) can significantly improve both singular models and ensembles. Sample-based methods have shown promising results in the field of medical image segmentation \citep{jungo2020analyzing,jungo2019assessing,mehrtash2020confidence}.

Other approaches use OOD data to explicitly train an \emph{outlier detector} \citep{bevandic2019simultaneous,hendrycks2018deep,lee2018training}. However, as they require OOD detection to be a primary goal throughout the training process, they cannot be applied post-hoc to pre-trained models.

Methods that modify or make certain assumptions on the architecture or training procedure have shown good performance \citep{kohl2018probabilistic,NEURIPS2020_95f8d990,monteiro2020stochastic,fuchs2021practical}. For instance, \emph{self-supervision} losses provide valuable assessments for novelty \citep{pidhorskyi2018generative,golan2018deep,hendrycks2019using,pmlr-v143-gonzalez21a}. However, their applicability to widely-used segmentation frameworks -- which do not typically use self-supervision -- is limited.

\begin{figure}[ht]
\centering
\includegraphics[width=0.9\columnwidth]{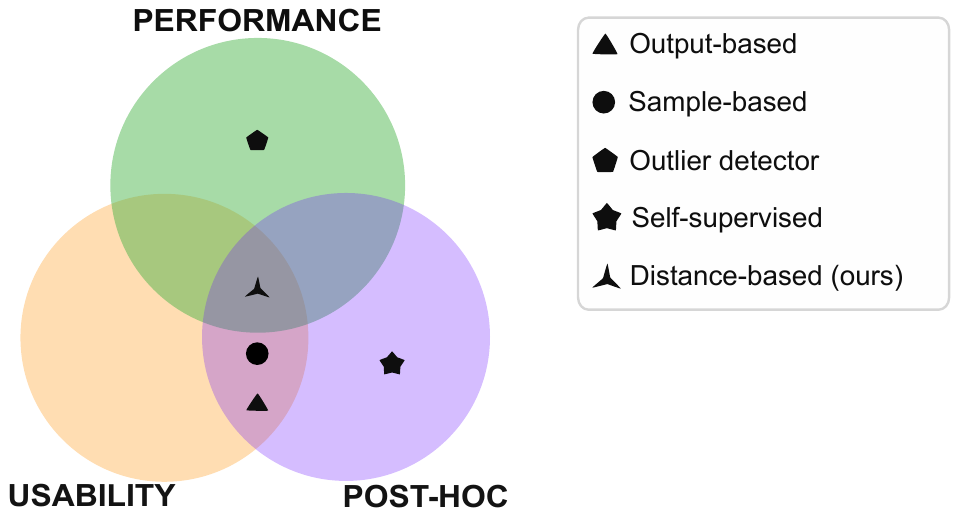}
\caption{\label{fig:OOD_methods} Desirable properties for OOD detection and corresponding paradigms. A method should ideally (1) be widely applicable (2) work on a \textit{post-hoc} basis even if OOD detection was not a goal during training and (3) reliably detect OOD samples.}
\end{figure}

In Figure \ref{fig:OOD_methods}, we illustrate how existing paradigms perform in terms of different desiderata. We are interested in approaches that can be directly used with any model, and so we restrict our analysis to the methods outlined in Table \ref{table:OOD_methods}.

\setlength{\tabcolsep}{3pt}
\begin{table}[h!]
\centering
\begin{adjustbox}{max width=\linewidth}
 \begin{tabular}{lcccc} 
\toprule
 \textbf{Method}    & \textbf{Type} & \textbf{Parameters} & \textbf{Mod. Level} & \textbf{Inf. time} \\ 
 \midrule
 Max. Softmax   & O     & t                 & 0 & ++ \\
 Temp. Scaling  & O     & t,T               & 1 & ++ \\
  KL   & O    & t, $p(\theta)$     & 2 & + \\
 Energy Scoring & O     & t,T               & 1 & ++ \\ 
 MC Dropout     & S    & t, p               & 3 & - \\
 TTA            & S    & t, $I_{Aug}$  & 2 & - - \\ 
 \textbf{Ours}           & D    & t, $\mu,\sigma$    & 2 & + \\ 
 \bottomrule
\end{tabular}
\end{adjustbox}
\caption{Comparison between Output- (O), Sample- (S) and Distance-based (D) methods. We compare important factors for applicability: parameters, number of modifications (0-3) and additional inference time from high [- -] to none [++].
}
\label{table:OOD_methods}
\end{table}

\begin{figure*}[!ht]
\centering
\includegraphics[width=0.9\textwidth]{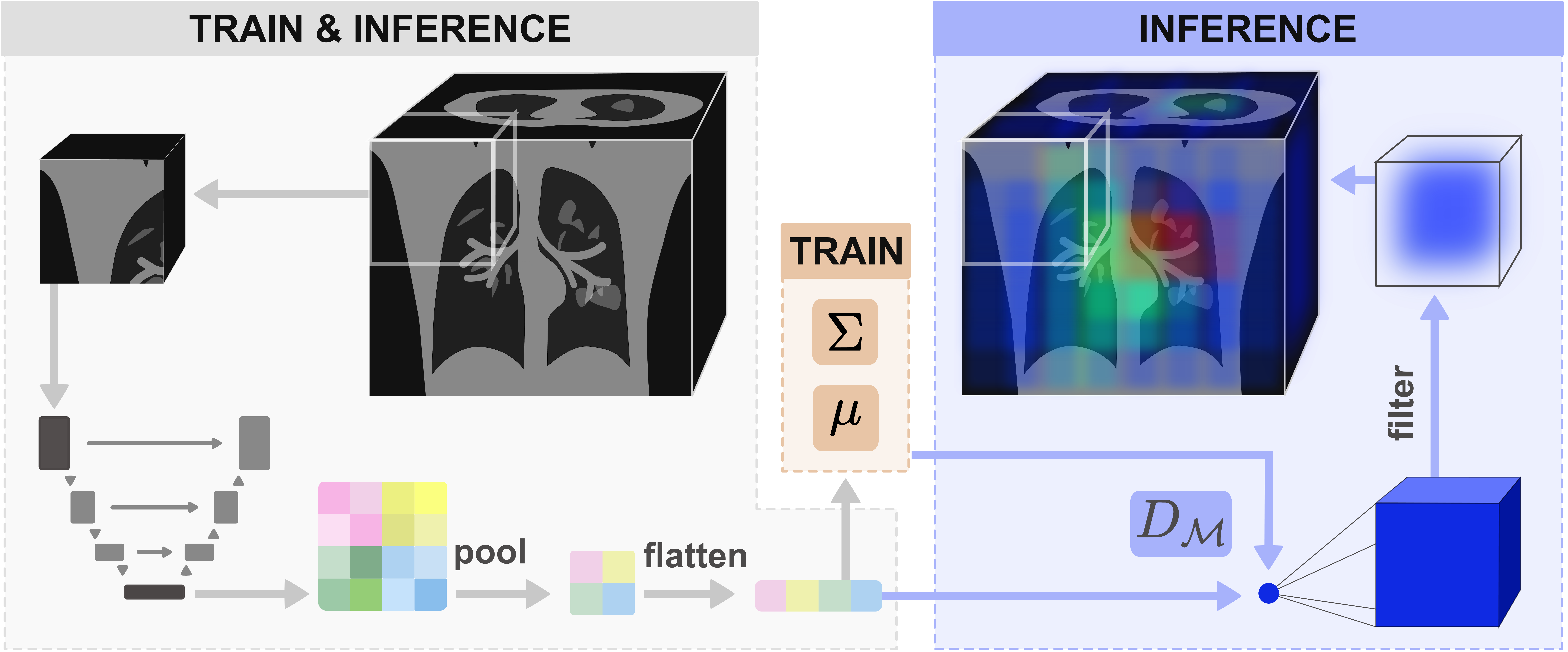}
\caption{Proposed method for OOD detection on a full-resolution nnU-Net model.} The input image first goes through a series of pre-processing steps and is divided into patches. For each patch, we take the feature maps generated at the end of the encoder during the forward pass. We then project these into a lower-dimensional, flattened subspace. During the training phase, we estimate a Gaussian distribution from the feature space by calculating $\mu$ and $\Sigma$. At inference time, we calculate the Mahalanobis distance to the training distribution and project the resulting point value into the dimensions of the original patch. Finally, a filtering operation is performed to weigh voxels at the centre more heavily, and the result is aggregated into a volume with the same dimensionality as the input image.
 \label{fig:methodology}
\end{figure*}

Unlike previous work, our method observes model activations at the end of the encoder. We project these to a lower-dimensional feature space and estimate a multi-variate Gaussian with the training data. During inference, we detect samples with a high \emph{Mahalanobis} distance to this distribution, which is suitable for quantifying differences in the latent space \citep{lee2018simple,ccalli2019frodo}.

\section{Material and methods}

Our proposed method, visualised in Figure \ref{fig:methodology}, assesses the uncertainty as the distance of new samples to the training distribution in the feature space. First, we extract feature maps from the trained model and project these to a low-dimensional space to ensure a computationally inexpensive calculation. We then estimate a multi-variate Gaussian distribution from ID train samples. At test time, we repeat the feature-extraction process and calculate the Mahalanobis distance. 

We first briefly introduce the patch-based nnU-Net architecture in Section~\ref{sec:nnunet} and outline how our method links to it. In Section~\ref{sec:est} we describe our proposed method for OOD detection, which follows a \emph{three-step process}: (1) estimation of a Gaussian distribution from training features (2) extraction of uncertainty masks for test images and finally (3) calculation of subject-level uncertainty scores.

\subsection{Patch-based nnU-Net} \label{sec:nnunet}

The nnU-Net is a standardised framework for medical image segmentation \citep{isensee2021nnu} that has reported state-of-the-art results across several benchmarks and challenges \citep{challenge_results}. Without deviating from the traditional U-Net structure \citep{ronneberger2015u}, it automatically chooses the best architecture and learning configuration for the training data. The framework also performs pre- and post-processing steps during both training and inference, such as adapting voxel spacing and normalising the intensities. 

We use the patch-based full-resolution variant, which is recommended for most applications \citep{isensee2021nnu}. After performing all necessary prepossessing operations, input image $x$ is divided into patches following a sliding window approach with an overlap of 50\%. This results in $N$ patches $\left\{ x_i\right\}^{N}_{i=1}$. A forward pass is made for each patch, at which point we extract feature maps for our method. Predictions for each patch are multiplied by a filtering operation that weights centre-voxels more heavily. Finally, weighted predictions are aggregated into an output mask with dimensionality of the original image. 

We also experiment with a 3D HighResNet model \citep{li2017compactness}, which we integrate into the nnU-Net framework and thus follow the same steps for image preparation and combination of the outputs into a coherent prediction.

\subsection{Distance-based OOD detection} \label{sec:est}

We are interested in capturing \textit{epistemic uncertainty}, which arises from a lack of knowledge about the data-generating process. While most uncertainty estimation methods quantify this uncertainty for prediction \textit{boundaries}, we want to do so for whole \textit{regions}, which is challenging for OOD data \citep{kendall2017uncertainties}.

One way to directly assess epistemic uncertainty is to calculate the distance between training and testing activations. As a model is unlikely to produce reasonable outputs for features far from any seen during training, this is a reliable signal for bad model performance \citep{lee2018simple}. 

\begin{figure}[ht]
\centering
\includegraphics[width=\columnwidth]{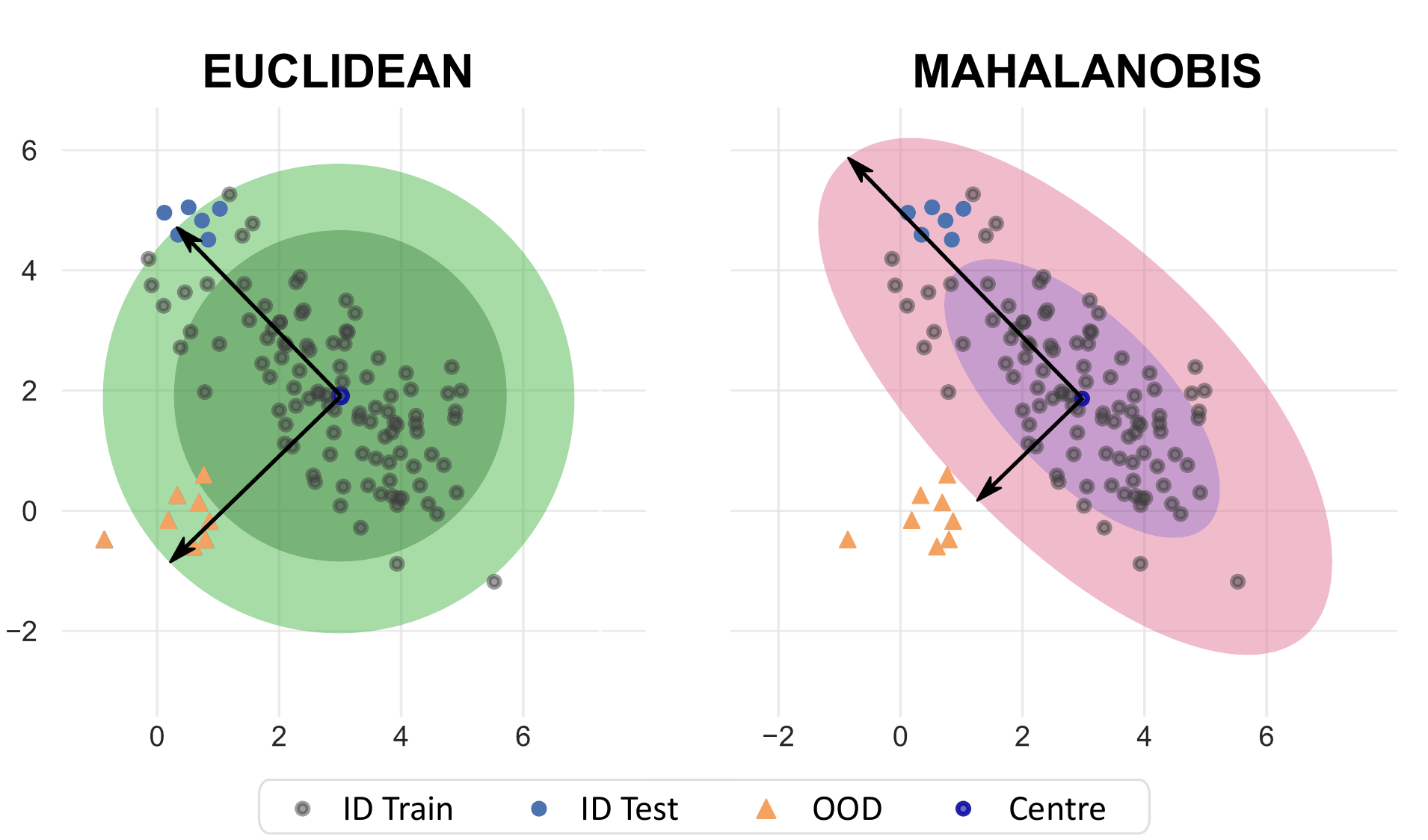}
\caption{Comparison between Euclidean and Mahalanobis distances in a two-dimensional space. Left: Euclidean distance fails to detect that OOD samples (orange triangles) strongly deviate from the expected behaviour of training samples (grey circles). Right: Mahalanobis distance adequately detects OOD samples, assigning them a distance outside the unit circle whilst properly admitting ID test samples (blue circles).} \label{fig:mahalanobis_merge}
\end{figure}

Model activations have covariance, and they do not necessarily resemble the mode for high-dimensional spaces \citep{wei2015understanding}, so the Euclidean distance is not appropriate for identifying unusual activation patterns. Instead, inspired by the work of \citet{lee2018simple}, we make use of the \emph{Mahalanobis distance} $D_\mathcal{M}$, which rescales samples into a space without covariance. Figure \ref{fig:mahalanobis_merge} illustrates how the Mahalanobis distance better captures the behaviour of in-distribution data and correctly identifies samples outside the unit circle as OOD.

The following sections describe how we leverage the Mahalanobis distance in our approach. Note that only one forward pass is necessary for each patch, keeping the computational overhead at a minimum.

 \subsubsection{Estimation of the training distribution} We start by estimating a multivariate Gaussian distribution $\mathcal{N}(\mu, \Sigma)$ over training features. For all training patches $\left\{ x_i\right\}^{N}_{i=1}$, features $\mathcal{F}(x_i)=z_i$ are extracted from the encoder $\mathcal{F}$. 
 
 For modern segmentation networks, the dimensionality of the extracted features $z_i$ is too large to calculate the covariance $\Sigma$ in an acceptable time frame. We thus project the latent space into a lower subspace by applying average Pooling operations with a kernel size of $\left ( 2,2,2 \right )$ and stride $\left ( 2,2,2 \right )$ until the dimensionality falls below $1e4$ elements. Finally, we flatten this subspace and estimate the empirical mean $\mu$ and covariance $\Sigma$. 

\begin{equation} \label{eq:cov}
    \mu = \frac{1}{N} \sum_{i=1}^N\hat{z_i}, \quad  \Sigma = \frac{1}{N} \sum_{i=1}^N (\hat{z_i} - \mu)(\hat{z_i} - \mu)^T
\end{equation}

 In Table \ref{tab:mahal_times} we demonstrate that for a dimensionality of $1e4$ elements we can estimate the covariance in a maximum of a few minutes (rows 3 and 4) with the \textit{Scikit Learn} on an AMD Ryzen 9 3900X CPU, whereas for higher dimensions the times increase abruptly (row 5).
\setlength{\tabcolsep}{4pt}
\begin{table}[ht]
\centering
\begin{adjustbox}{max width=\linewidth}
{\begin{tabular}{llll}
\toprule
Nr. samples & Dimensionality & $\Sigma$ time (s) & 
$D_\mathcal{M}$ time (s) \\
\midrule
1e3 & 1e3 & 0.260 & 0.001 \\
1e6 & 1e3 & 8.480 & 0.001 \\
1e3 & 1e4 & 69.11 & 0.050 \\
1e4 & 1e4 & 81.80 & 0.051 \\
1e3 & 2e4 & 6555.13 & 0.194 \\
\bottomrule
\end{tabular}}
\end{adjustbox}
\caption{Times in seconds required for estimating the covariance $\Sigma$ (column 3) and calculating the Mahalanobis distance $D_\mathcal{M}$ to one sample (column 4)}.
\label{tab:mahal_times}
\end{table}

\subsubsection{Extraction of uncertainty masks} During inference, we estimate an uncertainty mask for a subject following the process illustrated in Figure \ref{fig:methodology} (right). First, we perform the same preprocessing steps as during training and divide the image into patches. Next, we extract features maps for each patch $x_i$ and project them onto $\hat{z_i}$ as done during training. We then calculate the Mahalanobis distance (Eq.~\ref{eq:mahalanobis}) to the Gaussian distribution estimated in the previous step.

\begin{equation} \label{eq:mahalanobis}
    D_\mathcal{M}(\hat{z_i};\mu, \Sigma) = (\hat{z_i} - \mu)^T \Sigma^{-1}(\hat{z_i} - \mu)
\end{equation}

Each distance is a point estimate for the corresponding patch. We replicate this value to the size of the patch and combine the distances for all patches in the same manner as the segmentation pipeline combines patch outputs into a coherent prediction. 

Following the example of the patch-based nnU-Net, we start by initialising a zero-filled tensor with the dimensionality of the original image. We then apply a filtering operation to each patch to weigh voxels at the centre more heavily and add them to the image-level mask.

\subsubsection{Subject-level uncertainty} The previous step produces an uncertainty mask with the dimensionality of the input CT scan. In order to effectively identify highly uncertain images, we average over all voxels to obtain one value $\mathcal{U}$, and normalise uncertainties between the minimum and doubled maximum uncertainties for ID train data to ensure $\mathcal{U}\in \left[ 0, 1 \right]$.

\section{Experimental setup}

We start by describing the data used in our experiments in Section \ref{sec:data}. Afterwards, we state relevant details on our models (Section \ref{sec:model}). We then introduce all baselines (Section \ref{sec:baselines}) and define our evaluation metrics (Section \ref{sec:metrics}).

\subsection{Data} \label{sec:data}

 We train our first model with data from the \emph{COVID-19 Lung CT Lesion Segmentation Challenge} \citep{roth2021rapid,an2020ct,clark2013cancer}, which we refer to as \emph{Challenge} or in-distribution (ID). The dataset contains chest CT scans for patients with a confirmed SARS-CoV-2 infection from various centres and countries. The data is also heterogeneous in terms of age, gender, and disease severity of the patients. We use the 199 cases that are made available for the challenge, which we divide into 160 training and 39 testing cases with the nnU-Net random splitting function. 

We include results for four types of out-of-distribution samples: (1) \textbf{dataset shift}, where we evaluate the model on two other datasets with differences in the acquisition and population patterns (2) \textbf{transformation shift} where we apply artificial transformations to our ID data, (3) \textbf{diagnostic shift}, where we compare Covid-19 to non-Covid pneumonia patients, and (4) \textbf{far-OOD}, where we use the \textit{Spleen} and \textit{Colon} tasks of the Medical Segmentation Decathlon (MSD) \citep{simpson2019large,antonelli2022medical}.

In addition, we perform a study on hippocampus and prostate segmentation from MR images. We train each nnU-Net model with the corresponding task of the MSD and use two and three OOD datasets for hippocampus and prostate, respectively.

\subsubsection{Dataset shift}

We use two publicly available datasets: \textit{Mosmed} \citep{morozov2020mosmeddata} contains fifty cases and the \textit{Radiopedia} dataset \citep{ma_jun_2020_3757476}, a further twenty. Both encompass patients with and without confirmed infections. Table \ref{tab:data_chars} provides a summary of data characteristics.

\begin{table}[h!] 
\centering

\begin{adjustbox}{max width=\linewidth}
\begin{tabular}{llll}
\toprule
\textbf{Dataset name} & \textbf{Nr. cases} & \textbf{Mean image size} & \textbf{Mean spacing}\\
\midrule
Challenge & 199 & [512, 512, 69] & [0.8, 0.8, 4.8]\\
Mosmed & 50 & [512, 512, 41] & [0.7, 0.7, 8.0]\\
Radiopedia & 20 & [560, 571, 176] & [1.0, 1.0, 1.0]\\
\bottomrule
\end{tabular}
\end{adjustbox}
\caption{Characteristics of the Covid-19 lung lesion segmentation datasets.} \label{tab:data_chars}
\end{table}

\subsubsection{Transformation shift}

We transform the 39 in-distribution test cases with multiple operations from the \emph{TorchIO} \citep{perez-garcia_torchio_2021} library. 

\begin{table}[h!]
\centering
\begin{adjustbox}{max width=\linewidth}
 \begin{tabular}{l|llll} 
 \toprule 
   \textbf{Shift} & \textbf{Operation}  & \textbf{Weak} & \textbf{Medium} & \textbf{Strong} \\ 
 \midrule 
\multirow{4}{*}{Artefact}
  &Ghost intensity & (0, 0.2) & (0, 0.4) & (0, 0.7) \\ 
  &Spike intensity & (0, 0.2) & (0, 0.5) & (0, 0.7) \\ 
   &Blur STD &  (0, 0.3) &  (0, 0.3) &  (0, 0.3) \\ 
   &Noise STD &  (0, 15) &  (0, 30) &  (0, 30) \\ 
 \midrule 
 \multirow{4}{*}{Affine}&Scales & (0.9, 1.4) & (0.7, 1.8) & (0.6, 2) \\ [0.5ex] 
  &Rotation degrees & 5 & 8 & 9 \\ 
  &Translation range & (-15, 15) & (-20, 20) & (-20, 20) \\ 
  &Isotropic & True & True & False \\ 
   \bottomrule
\end{tabular}
\end{adjustbox}
\caption{ Parameters used to randomly generate artefacts and affine transformations with the \textit{TorchIO} library. For each type of shift, three transformed datasets are generated with increasingly stronger transformations.}
\label{table:augmentations:artifacts}
\end{table}

\begin{figure}[ht]
\centering
\includegraphics[width=0.85\columnwidth]{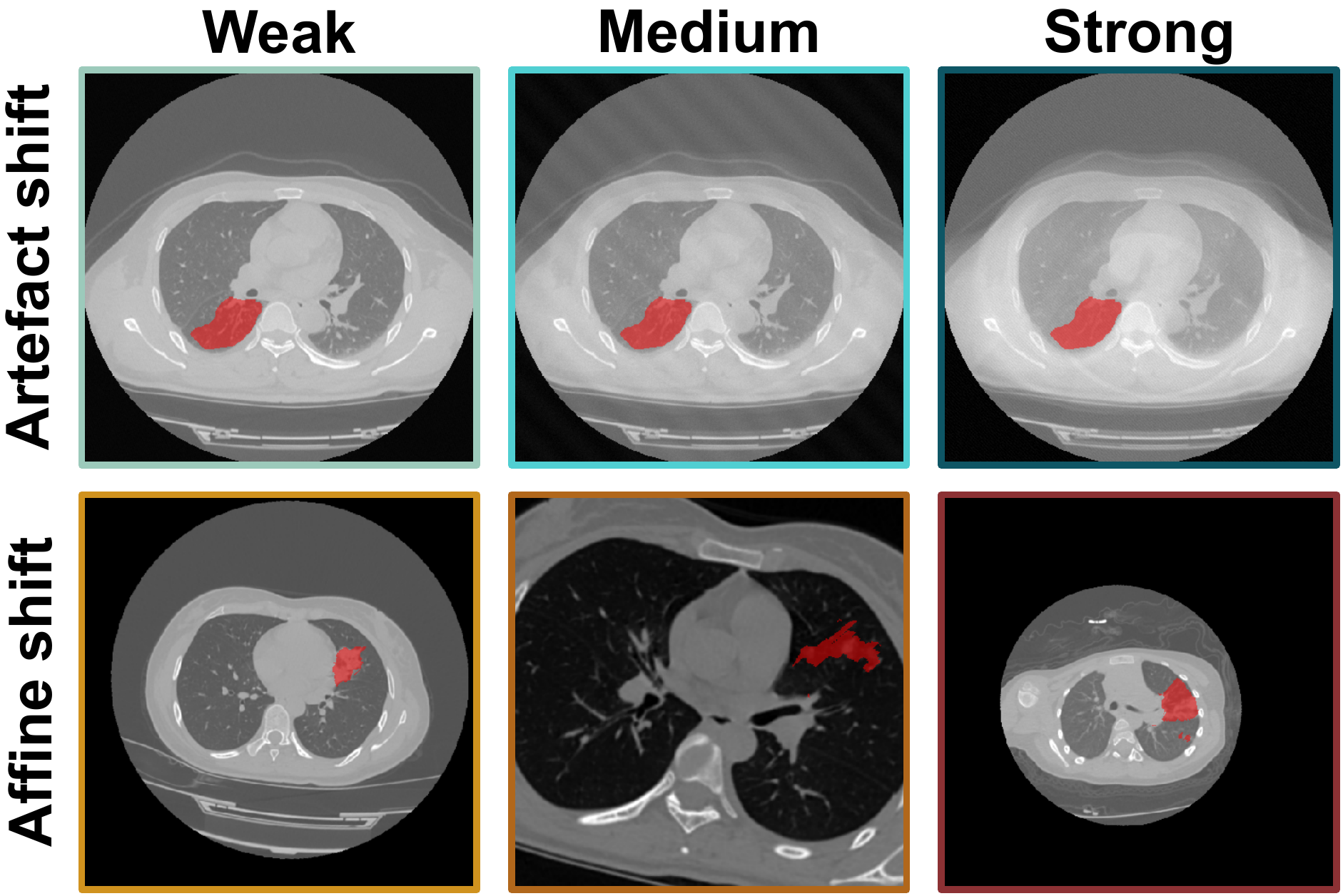}
\caption{Top row: Exemplary CT slice with overlaid segmentation mask in red after being transformed to contain artefacts in three magnitudes. Bottom row: Three exemplary CT slices with overlaid segmentation masks after applying affine transformations in three magnitudes. The border colours map each example to their corresponding datasets in Figure \ref{fig:boxplot_performance}.} \label{fig:augmentations}
\end{figure}

The \emph{artefact} transformations include ghosting, k-space spikes, Gaussian blurring, and Gaussian noise. \emph{Affine} transformations include scaling, rotation, and translation. All affine operations can be either isotropic or anisotropic. We deploy the same transformation parameters for the sagittal, coronal, and axial dimensions for the isotropic case. For the anisotropic case, these parameters change for every dimension, causing a stronger shift. For both groups of transformations, we generate three sets (\textit{weak, medium, and strong}), each with increasingly stronger augmentation parameters. The parameters used are reported in Table \ref{table:augmentations:artifacts}. Examples of the performed transformations are visualised in Figure \ref{fig:augmentations}.

\subsubsection{Diagnostic shift}

We utilise an in-house dataset of one hundred cases. Fifty patients have pulmonary infection of Covid-19 confirmed by RT PCR test and visible pulmonary Covid-19 lesions in all cases (3/2020 to 12/2020). The remaining fifty cases were composed of various Covid-mimics, manifesting similar pulmonary lesions but acquired prior to the Covid outbreak or tested negative for Covid-19 by RT PCR (3/2017 to 2/2020). Cases were collected and annotated in the RACOON project \citep{racoon}. Covid-mimics included are viral non-Covid pneumonia, bacterial pneumonia, fungal pneumonia, tuberculosis, chronic obstructive pulmonary disease, cystic fibrosis, interstitial pulmonary fibrosis, acute interstitial pneumonia, cryptogenic organising pneumonia, medication associated pulmonary toxicity, radiogenic pulmonary fibrosis, acute lung embolism, chronic lung embolism, pleural pathologies, pulmonary vasculitis, bronchial carcinoma, pulmonary metastasis, as well as a control case without any lung pathologies. 

A clinical radiologist with 8 years of experience in reading chest CT reviewed all scans and found them to be of good enough quality for accurate visual diagnosis. Manual annotations of the entire image stack were performed slice-by-slice by two independent readers trained in the delineation of GGOs and pulmonary consolidations. Central vascular structures and central bronchial structures were excluded from all annotations. Care was taken to differentiate between artefacts and GGO. Consolidations were defined as visible in a soft tissue window and at least 5 mm in size. An expert radiologist reader reviewed all delineations. In Table \ref{tab:in_house_data} we report some details on the demographic distribution.

\begin{table}[ht]
\centering
\begin{adjustbox}{max width=\linewidth}
{\begin{tabular}{lllll}
\toprule
& \textbf{Age} & \textbf{Gender} & \textbf{Voltage} & \textbf{mAs} \\
\midrule
Covid-19 & 57.17 [49/67] & 16\%& 100 & 121.21 $\pm$ 55.91\\
Non-Covid & 60.24 [47/73] & 42\% & 120 & 114.77 $\pm$ 82.56\\
\bottomrule
\end{tabular}}
\end{adjustbox}
\caption{In-house data cohort with 50 Covid-19 and 50 non-Covid cases. We report the age (median Q1/Q3), gender (f/m), voltage (median kV), and tube current-time product (mAs).}
\label{tab:in_house_data}
\end{table}

\subsubsection{MRI tasks}

For hippocampus we consider three T1-weighted datasets: the MSD task, which we denote \emph{MSD H}, and contains healthy and schizophrenia patients, 
the \emph{Dryad} \citep{kulaga2015multi} dataset with fifty healthy subjects 
and the \emph{Harmonized Hippocampal Protocol} data \citep{boccardi2015training} (\emph{HarP}) with senior subjects, some of which have Alzheimer’s.

For the segmentation of the prostate in T2-weighted MRIs we use a corpus of four datasets including the MSD data (\emph{MSD P}) and three OOD sets: the cases provided in the \textit{NCI-ISBI 2013 Challenge} \citep{isbi} (\emph{ISBI}) and the \emph{I2CVB} \citep{lemaitre2015computer} and \emph{UCL} \citep{litjens2014evaluation} datasets as made available by \citet{liu2020ms}.
To align label characteristics, we unify the labels of \textit{head} and \textit{body} for the hippocampus and of \textit{central gland} and \textit{peripheral area} for the prostate. A summary of the relevant dataset characteristics can be found in Table \ref{tab:data_hip_prostate_chars}.

\begin{table}[h!] 
\centering

\begin{adjustbox}{max width=\linewidth}
\begin{tabular}{llll}
\toprule
\textbf{Dataset name} & \textbf{Nr. cases} & \textbf{Mean image size} & \textbf{Mean spacing}\\
\midrule
MSD H & 260 & [50, 35, 36] & [1.0, 1.0, 1.0] \\
Dryad & 50 & [64, 64, 48] & [1.0, 1.0, 1.0] \\
HarP & 270 & [64, 64, 48] & [1.0, 1.0, 1.0] \\
\midrule
MSD P & 32 & [316, 316, 19] & [1.0, 1.0, 1.0] \\
ISBI & 30 & [384, 384, 19] & [0.5, 0.5, 3.7] \\
UCL & 13 & [384, 384, 24] & [0.5, 0.5, 3.3] \\
I2CVB & 19 & [384, 384, 64] & [0.5, 0.4, 1.3] \\
\bottomrule
\end{tabular}
\end{adjustbox}
\caption{Characteristics of the MR hippocampus (top) and prostate (bottom) segmentation datasets. Models were trained with the respective tasks of the \emph{Medical Segmentation Decathlon}.} \label{tab:data_hip_prostate_chars}
\end{table}

\subsection{Models} \label{sec:model}

We train three patch-based nnU-Nets \citep{isensee2021nnu} and one HighResNet \citep{li2017compactness} on a \textit{Tesla T4} GPU. Our configurations have patch sizes of $\left [ 256, 256, 28 \right ]$, $\left [ 56, 40, 40 \right ]$ and $\left [ 320, 320, 20 \right ]$ for the \textit{Challenge}, \textit{MSD H} and \textit{MSD P} tasks, respectively. In all cases, adjacent patches overlap by 50\%, and we train with a loss of Dice (smoothing 1e-5) and Binary Cross-entropy weighted equally until after convergence. Training begins with a learning rate of 0.01 and a weight decay of 3e-5. No test-time augmentation was applied to extract predictions, as this signifies a speed-up of 8 times for 3D data.

\subsection{Baselines} \label{sec:baselines}

We compare our approach to output- and sample-based techniques that assess uncertainty information by performing inference on a trained model. \emph{Max. Softmax} consists of taking the maximum softmax output \citep{hendrycks2016baseline}. \emph{Temp. Scaling} performs temperature scaling on the outputs before applying the softmax operation \citep{guo2017calibration}. \emph{KL from Uniform} computes the KL divergence from a uniform distribution \citep{hendrycks2019using}. Note that all three methods output a \textit{confidence} score (higher is more certain), which we invert to obtain an \textit{uncertainty} estimate (lower is more certain). \emph{Energy Scoring} \citep{liu2020energy} assesses uncertainty as the logarithmic sum of the softmax denominator. 

\emph{MC Dropout} ~\citep{gal2016dropout} consists of doing several forward passes whilst activating the Dropout layers that would usually be dormant during inference. We perform $10$ forward passes. Test-Time Augmentation (TTA) follows a similar strategy by augmenting images during testing \citep{wang2019aleatoric}. We use image-flip as augmentation and generate eight predictions by flipping the input image once clockwise and counter-clockwise for every axis. We report the standard deviation between outputs as an uncertainty score for both methods.

For all baselines and our proposed method we calculate a subject-level metric by averaging voxel values, and normalise the uncertainty range between the minimum and doubled maximum uncertainty represented in ID train data. For \emph{Energy Scoring} and \emph{Temp. Scaling}, we always report the result with lowest ESCE from among three different temperature settings $T \in \left \{ 1, 10, 100 \right \}$.

\subsection{Metrics} \label{sec:metrics}

For OOD detection, we calculate the 95\% \textit{true positive rate} (TPR) boundary on ID data, i.e. the boundary that covers at least 95\% of train samples. Samples with uncertainties greater than this boundary are predicted to be OOD. We report the \emph{false positive rate}, defined as

\begin{equation} \label{eq:fpr}
    FPR=\frac{FP}{FP+TN},
\end{equation}
where a \emph{false positive} (FP) is an OOD sample incorrectly deemed to be in-distribution, the \emph{Detection Error}
    \begin{equation} \label{eq:detection_error}
    Error=\frac{1}{2} (1-TPR)+\frac{1}{2} FPR
    \end{equation}
and the \emph{area under the receiving operating curve} (AUC), calculated with the \emph{Scikit Learn} library \citep{scikit_learn}.
    
While the detection of OOD samples is a first step in assessing the suitability of a model for a new image, an ideal uncertainty metric would inversely correlate with model performance. For this, we calculate the \textit{Expected Segmentation Calibration Error} (ESCE). Inspired by \citet{guo2017calibration}, we divide the $n$ test scans into $M=10$ interval bins $B_m$. For each bin, the absolute difference is calculated between average Dice $(Dice(B_m))$ and inverse average uncertainty $(1 - \mathcal{U}(B_m))$ for samples in the bin. A weighted average is reported that weights the score for each bin by the number of samples in it (Eq.~\ref{eq:ece}).

\begin{equation} \label{eq:ece}
    ESCE=\sum ^M_{m=1}\frac{\left | B_m \right |}{n}\left | Dice(B_m)-(1-\mathcal{U}(B_m)) \right |
\end{equation}

\section{Results}

\begin{figure*}[ht]
\centering
\includegraphics[width=\textwidth]{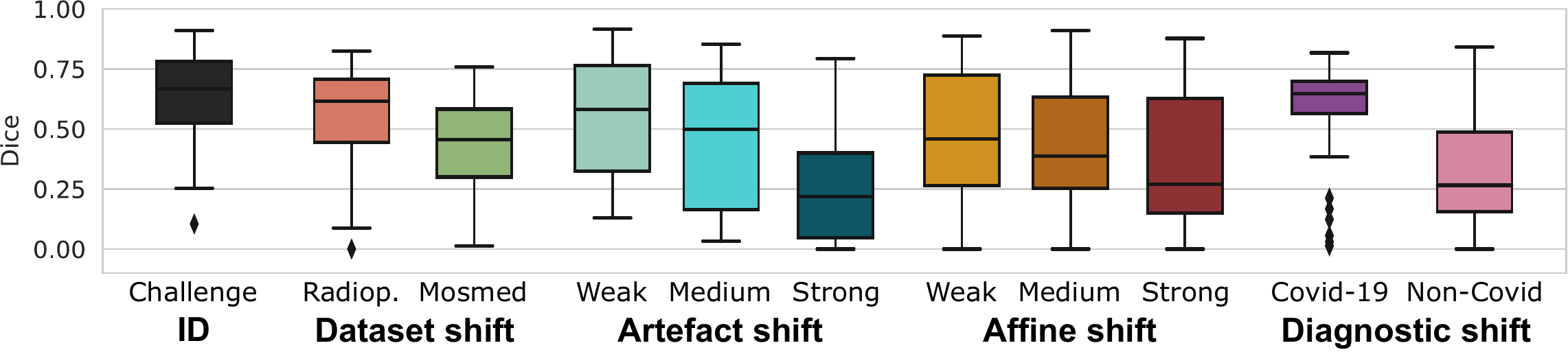}
\caption{Performance deterioration of a model trained with ID (\emph{Challenge}) data and tested on (1) \emph{Radiopedia} and \emph{Mosmed}; \emph{Challenge} test cases after applying  (2) \emph{artefact} and (3) \emph{affine} transformations with different levels of intensity; and (4) in-house Covid-19 and non-Covid pneumonia patients.} \label{fig:boxplot_performance}
\end{figure*}

We first analyse the \emph{dataset shift} scenario, where a model trained on the \emph{Challenge} dataset is tested on publicly available \emph{Radiopedia} and \emph{Mosmed} cases (Section \ref{sec:dataset_shift}). Afterwards, we evaluate how robust the model is against the presence of artefacts and affine transformations of different magnitudes and explore to what extent these are correctly detected (Section \ref{sec:transform_shifts}). As a third setting, we apply our method to an in-house data cohort with both Covid-19 and non-Covid patients in Section \ref{sec:diagnostic_shift}. 

In Section \ref{sec:farood}, we perform a \textit{far-OOD} study where we examine whether our method detects samples very far from the raining distribution. We then carry out an ablation study where we measure the use of different network layers for feature extraction (Section \ref{sec:ablation_study}) and repeat the \emph{dataset shift} experiments on a HighResNet model (Section \ref{sec:highresnet}). In all these experiments, we explore whether our method can distinguish between ID cases -- test subjects from the \emph{Challenge} data -- and OOD images. We qualitatively look into exemplary predictions and corresponding uncertainty scores in Section \ref{sec:qual_eval}.

Finally, in Section \ref{sec:results_mri}, we evaluate the transferability of our method to MR data, where we look at hippocampus and prostate segmentation tasks.

\subsection{Dataset shift} \label{sec:dataset_shift}

In Table \ref{tab:results_shifting_datasets}, we report the performance of our proposed method and six other approaches in identifying the OOD samples, i.e. samples from the \emph{Mosmed} or \emph{Radiopedia} datasets for which the model produces unreliable predictions (see Figure \ref{fig:boxplot_performance}). Following previous research in OOD detection \citep{liang2018enhancing}, we find the uncertainty boundary that covers 95\% of in-distribution train samples and deem cases with uncertainties beyond the ID 95th percentile threshold as OOD. Our distance-based method is the only approach that successfully flags cases far from the training distribution, as shown by a low detection error and FPR and an AUC close to one.

\setlength{\tabcolsep}{3pt}
\begin{table}[ht]
\centering
\begin{adjustbox}{max width=\linewidth}
{\begin{tabular}{lllll}
\toprule
\textbf{\textbf{Method}} & \textbf{ESCE $\downarrow$} & \textbf{Error $\downarrow$} & \textbf{FPR $\downarrow$} & \textbf{AUC $\uparrow$}\\
\midrule
Max. Softmax & .39 & .43 & .84 & .61\\
MC Dropout & .28 & .41 & .79 & .75\\
KL & .38 & .44 & .83 & .69\\
TTA & .36 & .41 & .77 & .74\\
Temp. Scaling & \textbf{.02} & .47 & .89 & .42\\
Energy Scoring & .46 & .51 & .90 & .31\\
\textbf{Ours} & .15 & \textbf{.09} & \textbf{.04} & \textbf{.96}\\
\bottomrule
\end{tabular}}
\end{adjustbox}
\caption{Dataset shift results. Ability of assessing segmentation quality as Estimated Segmentation Calibration Error (\emph{ESCE}) and identifying samples from \emph{Radiopedia} and \emph{Mosmed} as OOD in terms of Detection Error (\emph{Error}), False Positive Rate (\emph{FPR}) and Area Under the ROC (\emph{AUC}).}
\label{tab:results_shifting_datasets}
\end{table}

\begin{figure*}[h]
\centering
\includegraphics[width=\textwidth]{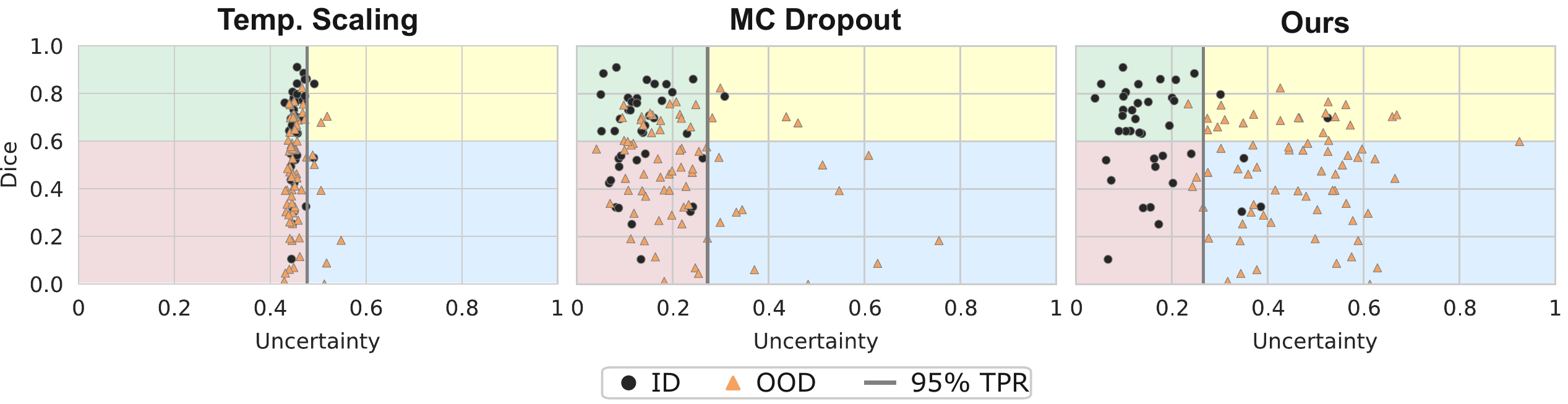}
\caption{Dice coefficient against normalised uncertainty for test ID (black circles) and OOD (orange triangles) scans. The ID samples are from the \emph{Challenge} dataset, and the OOD ones from \emph{Mosmed} or \emph{Radiopedia}. The grey vertical line marks the 95\% TPR for ID train data. Samples to the right are predicted to be OOD. Clinically relevant is the lower left (red) quadrant that houses silent failures, i.e. predictions with a Dice $<$ 0.6 and low uncertainty scores.} \label{fig:scatters}
\end{figure*}

We plot the Dice score against normalised uncertainty for the three best-performing methods in Figure \ref{fig:scatters}. The vertical line marks the 95\% TPR boundary. We consider predictions with a Dice score lower than 0.6 to be of \emph{low quality} as they diverge significantly from the ground truth \citep{valindria2017reverse} and, for the task of Covid-19 lesion segmentation, provide a misleading assessment of the spread of the infection.

The lower left (red) quadrant is critical for the safe use of segmentation models, as it houses \emph{silent failures} for which \emph{low-quality predictions} are made but which are not identified as such. Only our method assigns sufficiently large uncertainty estimates to poorly segmented OOD samples, excluding them from this section. Nevertheless, the upper right (yellow) quadrant shows that our method is too conservative in estimating uncertainties, not identifying samples for which the model produces good segmentations. This overly cautious behaviour potentially leads to an under-utilisation of the model for cases that are technically OOD but have very apparent lesions which are easy to segment; though any amount of safe utilisation is advantageous. Another limitation of the proposed method is that it fails to identify ID samples that the model segments incorrectly due to the lesions being too small or different from those seen in the training data, highlighting the fact that OOD detection is only part of a thorough QA process.

Regarding the estimation of segmentation quality, \emph{Temp. Scaling} reaches the lowest ESCE (first column in Table \ref{tab:results_shifting_datasets}), but a closer inspection of Figure \ref{fig:scatters} (left) displays that this is due to most uncertainties clustering on the fifth bin. An ideal segmentation calibration would house all samples in the upper left (green) and lower right (blue) quadrants.

\subsection{Artefact and affine shifts} \label{sec:transform_shifts}

The \emph{dataset shift} scenario observed in the previous section depicts a realistic setting whether there are several potential degrees of variation between the training data and cases encountered during deployment. However, it is difficult to assess whether the model performance falls due to (a) changes in the acquisition process, (b) another patient population or simply (c) a different delineation process for ground truth segmentation masks. Subsequently, we cannot confidently assess \emph{why} cases are flagged as OOD. We therefore artificially transform the same ID test cases in two different ways and three levels of magnitude. More than any other explored scenario, these images could be deemed  \textit{near-OOD} \citep{fort2021exploring}. Nevertheless, there is a significant performance deterioration for transformed images, which grows with the magnitude of the perturbation (Figure \ref{fig:boxplot_performance}).

\begin{figure}[h!]
\centering
\includegraphics[width=\columnwidth]{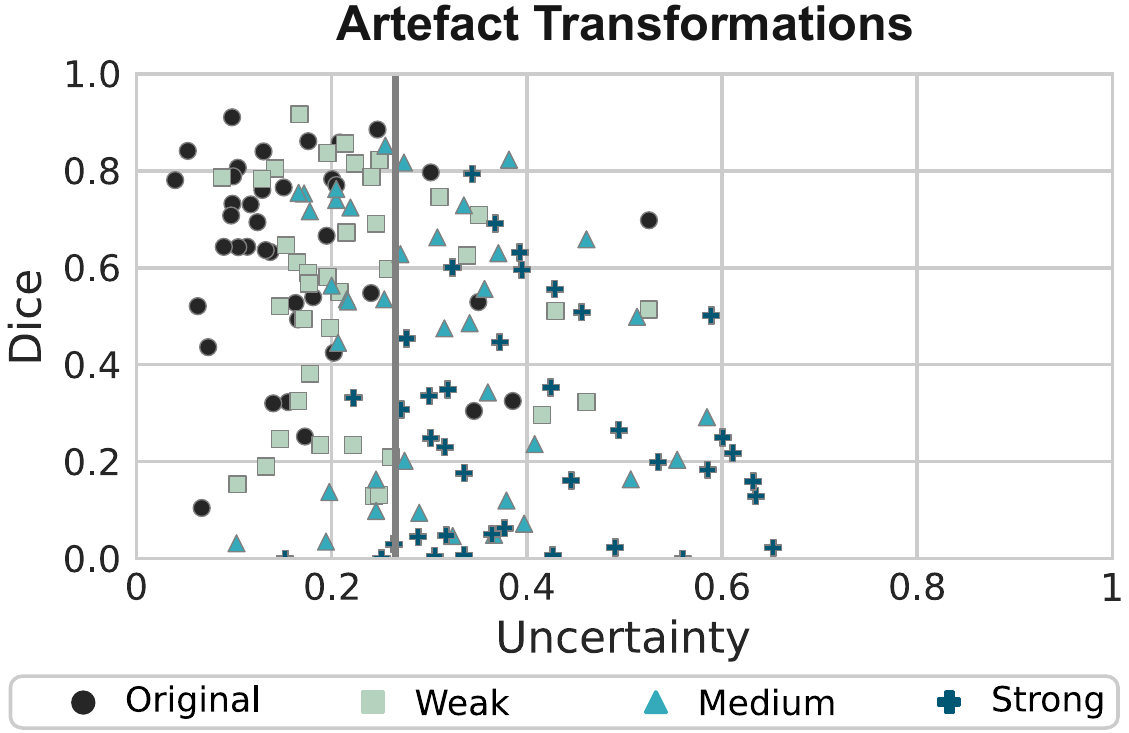}
\caption{Dice coefficient against normalised uncertainty. Black circles are the test ID (unmodified \emph{Challenge}) images, and the remaining markers stand for the same \emph{Challenge} images after applying transformations to simulate common artefacts.} \label{fig:scatter_artifacts}
\end{figure}

We start by simulating the presence of common image artefacts. In Figure \ref{fig:scatter_artifacts}, we visualise the results of our method.

While non-transformed (\emph{original}) cases are correctly assigned low uncertainty scores and most heavily transformed samples are identified as OOD, several samples for which bad segmentations are produced are not identified. Most of these are only weakly transformed (mint-coloured squares). On the other hand, many weakly transformed cases for which good segmentations are produced are correctly assigned low uncertainties despite not being ID. Most heavily transformed images (turquoise crosses) are correctly deemed too far from the training distribution to have reliable predictions.

\begin{figure*}
\centering
\includegraphics[width=\textwidth]{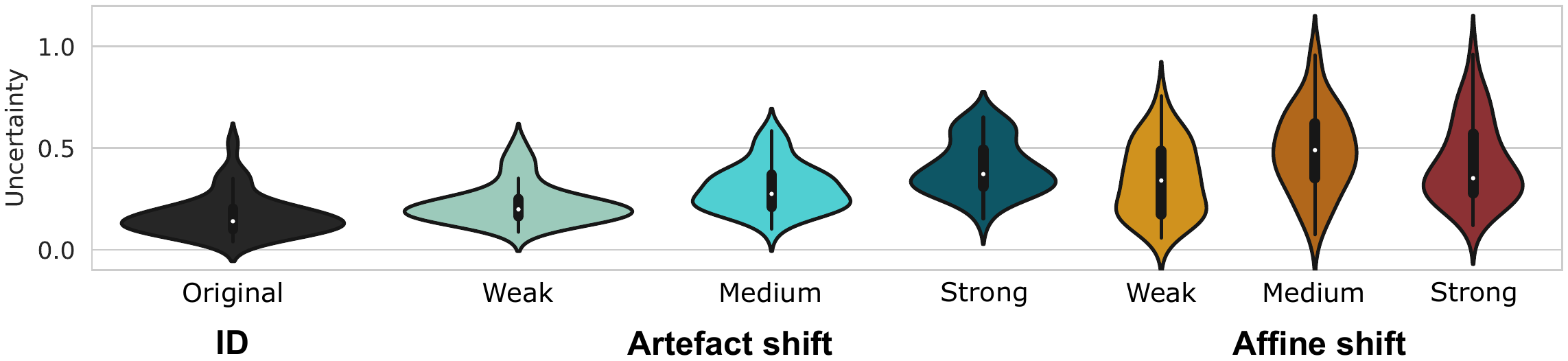}
\caption{Distribution of uncertainty scores estimated by our proposed method for the \emph{artefact shift} and \emph{affine shift} scenarios. In general, the uncertainties increase with the intensity of the transformations.} \label{fig:violin_transformed}
\end{figure*}

\begin{figure}[h!]
\centering
\includegraphics[width=\columnwidth]{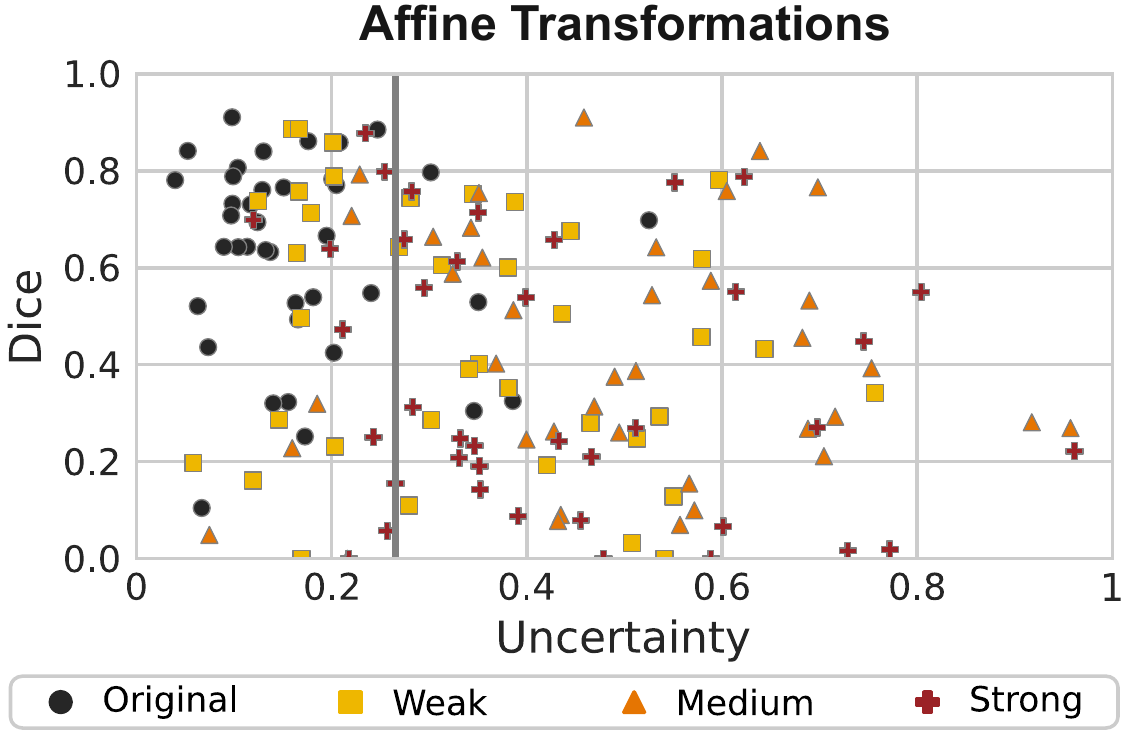}
\caption{Dice coefficient against normalised uncertainty. Black circles are the test ID (unmodified \emph{Challenge}) images, and the remaining markers stand for the same \emph{Challenge} images after applying transformations to simulate affine shifts.} \label{fig:scatter_affine}
\end{figure}

A similar situation occurs when we apply affine transformations to simulate geometric changes (Figure \ref{fig:scatter_affine}). These could arise from shifting population patterns, scans being acquired for different ranges, or using other acquisition parameters. Our method deems many weakly transformed cases (yellow squares) to be ID. This is positive as good segmentations are available for most cases. However, a few failure cases are not adequately identified.

Table \ref{tab:results_transforms} compares several approaches in terms of OOD detection and segmentation quality assessment. While our method displays an acceptable calibration error and the best OOD detection performance, this \emph{near-OOD} problem proves more difficult than \emph{dataset shift}. It particularly seems to be very difficult to reliably detect image artefacts.

\begin{table}[ht]
\centering
\begin{adjustbox}{max width=\linewidth}
{\begin{tabular}{lllll}
\toprule
\textbf{\textbf{Method}} & \textbf{ESCE $\downarrow$} & \textbf{Error $\downarrow$} & \textbf{FPR $\downarrow$} & \textbf{AUC $\uparrow$}\\
\midrule
Max. Softmax & .46/.44 & .48/.46 & .94/.89 & .55/.56\\
MC Dropout & .44/.44 & .51/.51 & 1.0/.99 & .22/.23\\
KL & .46/.44 & .48/.46 & .91/.86 & .58/.57\\
TTA & .43/.41 & .46/.38 & .87/.72 & .63/.61\\
Temp. Scaling & \textbf{.05/.04} & .51/.35 & .95/.62 & .50/.76\\
Energy Scoring & .52/.51 & .53/.33 & .92/.53 & .49/.76\\
\textbf{Ours} & .26/.21 & \textbf{.29/.18} & \textbf{.45/.24} & \textbf{.83/.89}\\
\bottomrule
\end{tabular}}
\end{adjustbox}
\caption{Transformation shift results. Segmentation calibration (as ESCE) and OOD detection scores between original \emph{Challenge} images and cases modified with synthetic artefacts and affine transformations, respectively.}
\label{tab:results_transforms}
\end{table}

We further visualise the uncertainty ranges assigned to each shift and magnitude in Figure \ref{fig:violin_transformed}. As expected, the uncertainty increases with the degree of transformation for artefact shifts. For affine shifts, \emph{medium} changes result in similar uncertainties to \emph{strong} ones. This is likely due to the selected transformation sequences being too similar (see Table \ref{table:augmentations:artifacts}), which results in a similar performance for \emph{medium} and \emph{strong} artefacts (Figure \ref{fig:boxplot_performance}).

In general, we can conclude that the uncertainty correlates positively with the degree of deformation and inversely with model performance. Affine transformations also have a more pronounced effect on the uncertainties (Figure \ref{fig:violin_transformed}). This possibly stems from the training data containing similar patterns to those introduced by the weaker artefact transformations.

\subsection{Diagnostic shift} \label{sec:diagnostic_shift}

\begin{figure}[h!]
\centering
\includegraphics[width=\columnwidth]{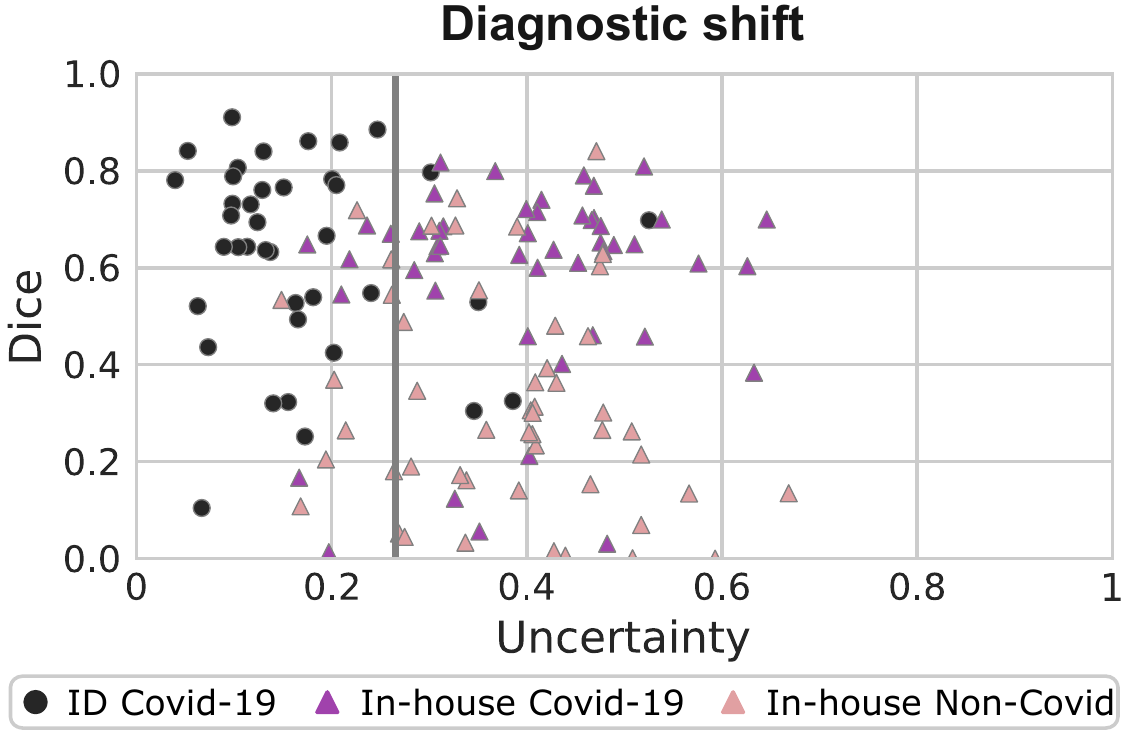}
\caption{Dice coefficient against normalised uncertainty for ID test (\emph{Challenge})} data and in-house \emph{chest CTs} of Covid-19-positive (purple triangles) and non-Covid (pink triangles) patients. \label{fig:scatter_inhouse}
\end{figure}

We have not yet analysed how the segmentation model performs across disease patterns. To explore this, we segment lung lesions in the form of GGOs and consolidations for an in-house cohort of 50 Covid-19 and 50 non-Covid cases. The performance of the model on the non-Covid cases is significantly worse. Table \ref{tab:results_in_house} summarises our findings, and we plot our uncertainty assessment in Figure \ref{fig:scatter_inhouse}. 

\begin{table}[h!]
\centering
\begin{adjustbox}{max width=\linewidth}
{\begin{tabular}{lllll}
\toprule
\textbf{\textbf{Method}} & \textbf{ESCE $\downarrow$} & \textbf{Error $\downarrow$} & \textbf{FPR $\downarrow$} & \textbf{AUC $\uparrow$}\\
\midrule
Max. Softmax & .29/.42 & .22/.32  & .42/.62 & .86/.87\\
MC Dropout & .22/.38 & .30/.46 & .58/.90 & .84/.69\\
KL & .29/.42 & .23/.33 & .40/.60 & .88/.89\\
TTA & .25/.32 & .19/.17 & .32/.28 & .89/\textbf{.95}\\
Temp. Scaling & \textbf{.07/.05} & .34/.54 & .62/1.0 & .78/.06\\
Energy Scoring & .38/.54 & .49/.56 & .86/1.0 & .61/.05\\
\textbf{Ours} & .16/.26 & \textbf{.13/.15} & \textbf{.14/.18} & \textbf{.93}/.92 \\
\bottomrule
\end{tabular}}
\end{adjustbox}
\caption{Diagnostic shift results. Segmentation calibration (as ESCE) and OOD detection scores between test ID \emph{Challenge} images and in-house cases with and without Covid-19, respectively.}
\label{tab:results_in_house}
\end{table}

Our method reliably detects cases from our in-house cohort, though it does not distinguish between Covid-19 and non-Covid cases. Though ideally Covid-19 cases for which good predictions are produced should be deemed low-uncertainty, the fact that badly segmented non-Covid cases are flagged as OOD is more relevant for clinical use as unsure good predictions are preferred over confident faulty ones.

\subsection{Far-OOD examinations} \label{sec:farood}

We have extensively examined \emph{near-OOD} \citep{fort2021exploring} cases where a performance deterioration is unexpected. In contrast, \emph{far-OOD} situations occur when an input is erroneously fed into a model, and there is no realistic expectation that a model can produce a sensible prediction.

In Table \ref{tab:results_farOOD}, we examine what happens when we feed CT spleen and colon cancer examinations from the \textit{Medical Segmentation Decathlon} into our model trained to segment pulmonary lesions from chest CTs. Our method  distinguishes between ID and far-OOD cases, correctly identifying all colon examinations as OOD (FPR = 0) and showing detection errors of up to 0.1 for both anatomies.

\begin{table}
\centering
\begin{adjustbox}{max width=\linewidth}
{\begin{tabular}{lllll}
\toprule
\textbf{\textbf{Method}} & \textbf{ESCE $\downarrow$} & \textbf{Error $\downarrow$} & \textbf{FPR $\downarrow$} & \textbf{AUC $\uparrow$}\\
\midrule
Max. Softmax & .58/.71 & .44/.42  & .85/.81 & .89/.89\\
MC Dropout & .50/.64 & .37/.36 & .68/.66 & .88/.87\\
KL & .59/.72 & .44/.42 & .85/.81 & .88/.88\\
TTA & .48/.58 & .18/.22 & .29/.37 & .95/.95\\
Temp. Scaling & .62/.71 & .48/.42 & .93/.81 & .79/.89\\
Energy Scoring & \textbf{.31}/\textbf{.16} & .49/.51 & .93/1.0 & .50/.50\\
\textbf{Ours} & .34/.41 & \textbf{.10}/\textbf{.06} & \textbf{.07}/\textbf{.00} & \textbf{.96}/\textbf{.98} \\
\bottomrule
\end{tabular}}
\end{adjustbox}
\caption{Far-OOD results. Segmentation calibration (as ESCE) and OOD detection scores between test ID \emph{Challenge} images and CT scans for spleen and colon examinations, respectively.}
\label{tab:results_farOOD}
\end{table}

\subsection{Ablation study} \label{sec:ablation_study}

We evaluate which features are most expressive for detecting distribution shifts in Table \ref{tab:ablation}. We compare the use of activations at the middle of the network, more specifically the convolutional (Conv) parameters of the sixth \emph{encoding block} (EB) against those of the first \emph{decoding block} (DB), and features at the beginning (1st EB) and final end (6th DB) of the architecture. In addition, we look into the use of \emph{batch normalisation} (BN) layers, as these normalise layer inputs and therefore contain domain information \citep{ioffe2015batch}. The results show that features at the middle of the network (\emph{6th EB Conv}, followed by \emph{6th EB BN} and \emph{1st DB Conv}) are the most suitable for detecting distribution shifts.

\begin{table}[ht]
\centering
\begin{adjustbox}{max width=\linewidth}
{\begin{tabular}{lllll}
\toprule
\textbf{Features} & \textbf{ESCE} $\downarrow$ & \textbf{Error} $\downarrow$ & \textbf{FPR} $\downarrow$ & \textbf{AUC} $\uparrow$ \\
\midrule
\textbf{6th EB Conv} & \textbf{.15/.23} & \textbf{.09/.24} & .04/.35 & \textbf{.96/.86} \\
6th EB BN & .18/.23 & .11/.25 & .09/.37 & .95/.85 \\
1st EB Conv & .42/.24 & .56/.70 & .13/.40 & .81/.21 \\
1st EB BN & .52/.45 & .50/.50 & \textbf{.00/.00} & .51/.51 \\
1st DB Conv & .17/.25 & .09/.25 & .06/.38 & \textbf{.96}/.84 \\
6th DB Conv & .52/.45  & .50/.50 & \textbf{.00/.00} & .50/.50 \\
\bottomrule
\end{tabular}}
\end{adjustbox}
\caption{Ablation study on the usability of feature maps. OOD detection and segmentation calibration for our proposed method using different convolutional (Conv) and batch normalisation (BN) at different encoding (EB) and decoding blocks (DB). The results are for the \emph{dataset shift} and \emph{transformed} (including both artefact and affine shifts) scenarios, respectively.}
\label{tab:ablation}
\end{table}

\subsection{HighResNet model} \label{sec:highresnet}

Not all segmentation models follow an encoder-decoder structure. For instance, the HighResNet \citep{li2017compactness} uses dilated convolutions and residual blocks to produce accurate segmentations. That raises the questions of whether our proposed approach would be effective on this architecture and which features would be most helpful for detecting distribution shifts. We report these results for the \textit{dataset shift} scenario in Table \ref{tab:highresnet}. The upper section summarises the results for all baselines, and the lower part shows the performance of our proposed method for three different feature maps.

\setlength{\tabcolsep}{3pt}
\begin{table}[ht]
\centering
\begin{adjustbox}{max width=\linewidth}
{\begin{tabular}{lllll}
\toprule
\textbf{\textbf{Method}} & \textbf{ESCE $\downarrow$} & \textbf{Error $\downarrow$} & \textbf{FPR $\downarrow$} & \textbf{AUC $\uparrow$}\\
\midrule
Max. Softmax & .35 & .48 & .94 & .57\\
MC Dropout & .35 & .49 & .96 & .59\\
KL & .34 & .46 & .90 & .60\\
TTA & .35 & .48 & .90 & .61\\
Temp. Scaling & .35 & .48 & .93 & .54\\
Energy Scoring & .58 & .49 & .97 & .50\\
\midrule
7th Conv Block & .41 & .47 & \textbf{.00} & \textbf{.94}\\
6th Dil Conv Block & .58 & .50 & \textbf{.00} & .50\\
\textbf{12th Dil Conv Block} & \textbf{.33} & \textbf{.37} & \textbf{.00} & .84\\
\bottomrule
\end{tabular}}
\end{adjustbox}
\caption{HighResNet results. Segmentation calibration (as ESCE) and OOD detection scores between test ID \emph{Challenge} images and OOD samples belonging to the \emph{Radiopedia} or \emph{Mosmed} datasets, for a HighResNet model trained on \emph{Challenge}. The bottom part of the table shows three variations of our method with different feature maps: the 7th conv. block, the 6th block with dilated conv., and the 12th (last) block with dilated convolutions.}
\label{tab:highresnet}
\end{table}

The HighResNet architecture is divided into four sections: (1) seven convolutional blocks, (2) six blocks with dilated convolutions using a dilation factor of 2, (3) six dilated convolutional blocks with a factor of 4, and (4) a final convolutional block. Residual connections with identity mapping are also included every two blocks to join features at different levels. We test the use of three feature maps: the last (7th) convolutional block, the last (6th) dilated convolutional block with factor 2, and the last (12th) dilated convolutional block.

\begin{figure}[h!]
\centering
\includegraphics[width=\columnwidth]{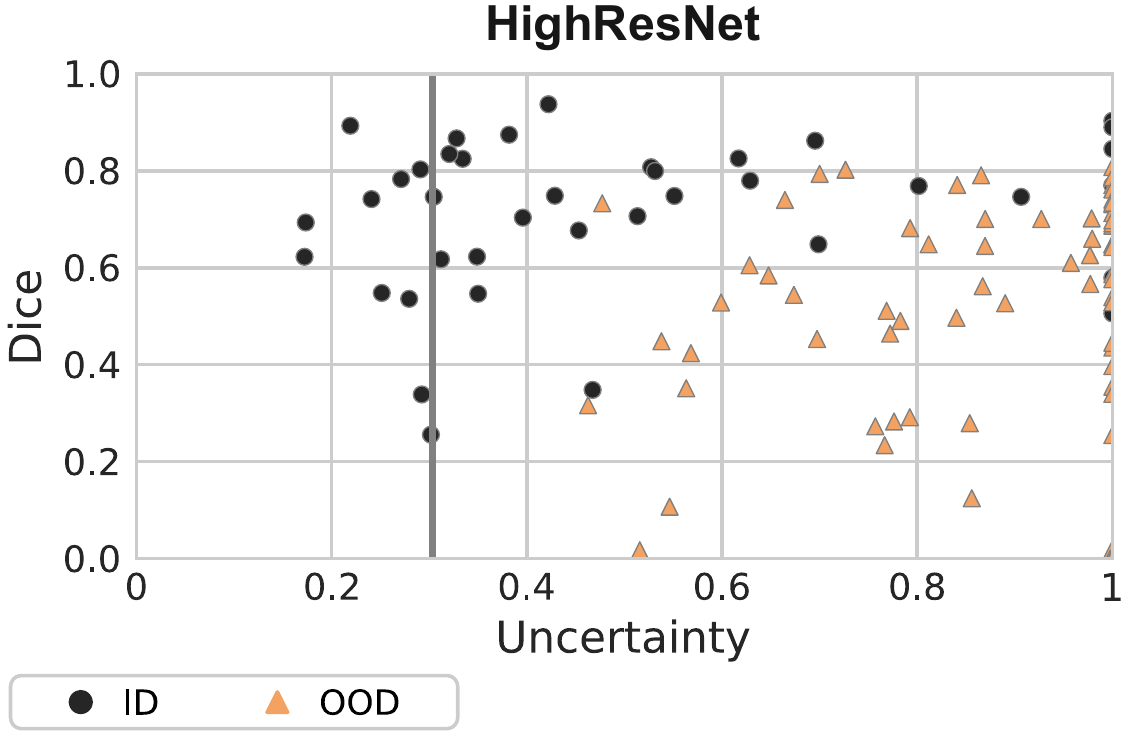}
\caption{Dice coefficient against normalised uncertainty for the variant using the 12th Dil. Conv. Block. Black circles are test ID (\emph{Challenge}) images, and orange triangles are OOD cases from \emph{Radiopedia} or \emph{Mosmed}.} \label{fig:highresnet}
\end{figure}

The best results are for the variant of our method which uses the last block with dilated convolutions. Though the FPR and AUC are encouraging, the detection error is relatively high, suggesting that the TPR is low as the 95\% TPR on ID train data does not cover a significant portion of ID test samples (see Eq. \ref{eq:detection_error}). We plot the performance of the network vs. normalised uncertainties for the best-performing features in Figure \ref{fig:highresnet}. A separation is noticeable between ID (Challenge) and OOD (Radiopedia and Mosmed), but the uncertainty boundary -- as hypothesised from the high Detection Error -- is too low. This means that OOD samples are correctly detected, yet the model is under-utilised.

\subsection{Qualitative evaluation} \label{sec:qual_eval}

\begin{figure*}
\centering
\includegraphics[width=0.9\textwidth]{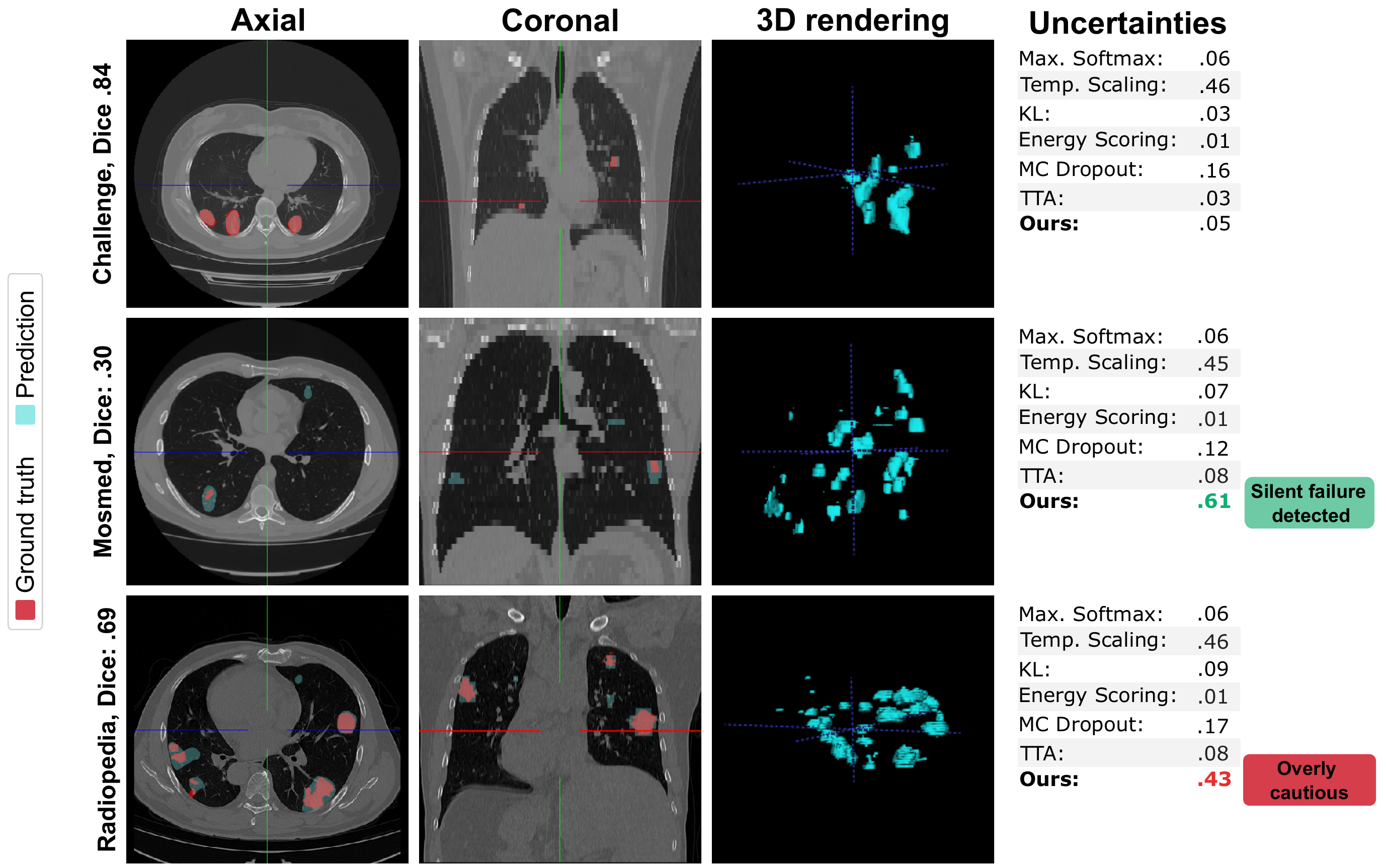}
\caption{Axial and coronal slices with overlaid predictions and ground truths and volume renderings of the predictions for three different subjects. First column: a good prediction. Second column: a poor prediction for an OOD case which our method successfully detects. Though there are considerable differences to the ground truth, these errors are not directly noticeable even for trained observers. Third column: a good prediction for an OOD case.} \label{fig:qual}
\end{figure*}

We now take a detailed view of some cases in Figure \ref{fig:qual}. The first column shows an in-distribution \emph{Challenge} case with a good prediction. The second and third cases are from \emph{Mosmed} and \emph{Radiopedia}, respectively. While the \emph{Mosmed} prediction is significantly different from the ground truth (incorrectly marking several regions as lesions), a good segmentation is produced for the third case.

We first notice the complexity of assessing whether a segmentation mask for lung lesions is correct. An untrained observer would not be able to detect that the second segmentation is so different from the ground truth, and even trained radiologists may not directly identify this error, as GGOs can manifest in superior lobes and with multiple connected components \citep{parekh2020review}. Similarly, all methods fail to detect this case except for our distance-based method, which assigns an uncertainty of 0.61.

The prediction for the third case over-segments some lesions, though if we observe the difference between the \emph{Challenge} and \emph{Radiopedia} ground truth masks, we notice that delineations are courser for the first case (we see in the first image that broad regions around lesions are marked as infected). Therefore, the model learns to mimic this behaviour. Beyond this, the segmentation model correctly detects all lesions and only creates a very small additional component. Here, our method makes an overly cautious uncertainty assessment, assigning this case an uncertainty of .43 which falls beyond the 95\% TPR boundary.

\subsection{Application to MRI data} \label{sec:results_mri}

Magnetic Resonance Imaging (MRI) data is even more susceptible to changes in the acquisition conditions than CTs, as there is no consensus on the calibration of intensity values. This causes the performance of segmentation models trained on MR tasks to deteriorate on OOD data \citep{zakazov2021anatomy,kondrateva2021domain}.

\begin{figure}[h!]
\centering
\includegraphics[width=\columnwidth]{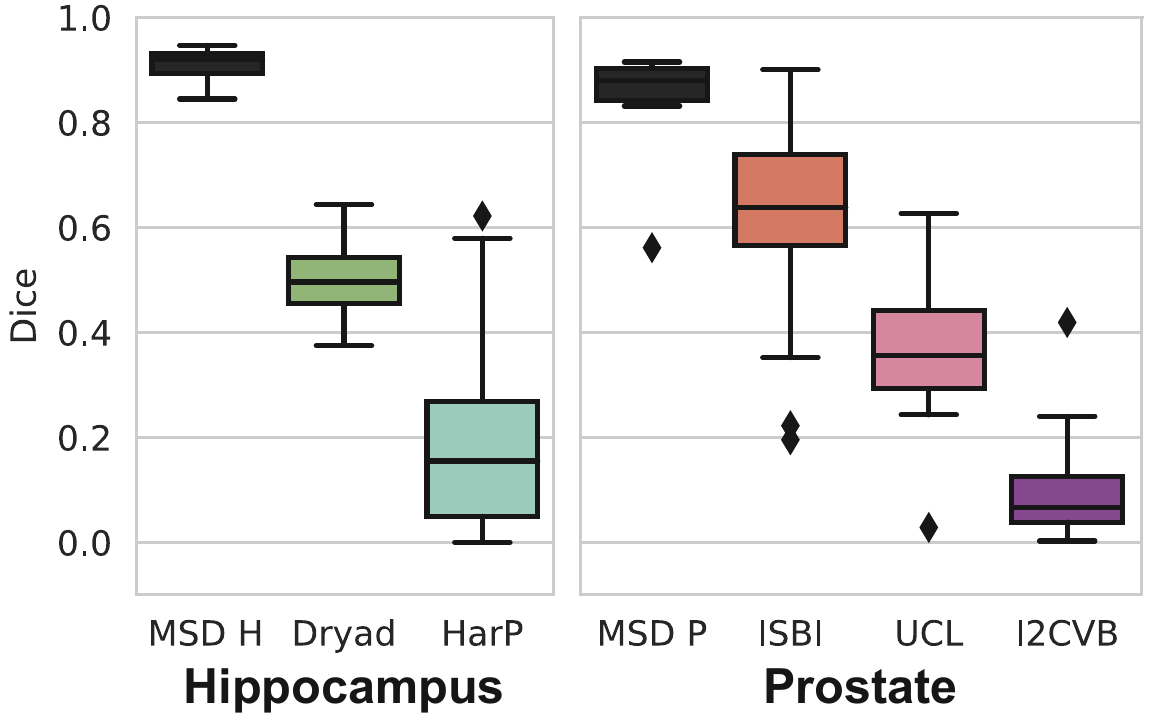}
\caption{Performance as Dice score of models trained with \emph{MSD H} (left) and \emph{MSD P} (right) data for hippocampus and prostate segmentation, respectively. Plotted are the ID test (in dark blue) and OOD scores.} \label{fig:prost_hip_performance}
\end{figure}

In this section, we evaluate how our proposed method can help detect such distribution shifts on nnU-Net models trained with the \textit{hippocampus} and \textit{prostate} tasks of the MSD. Figure \ref{fig:prost_hip_performance} illustrates that while the initial performance of the models is over 0.8 Dice on in-distribution test data (\emph{MSD H} and \emph{MSD P}), it falls significantly for the OOD datasets.

\begin{figure}[h!]
\centering
\includegraphics[width=\columnwidth]{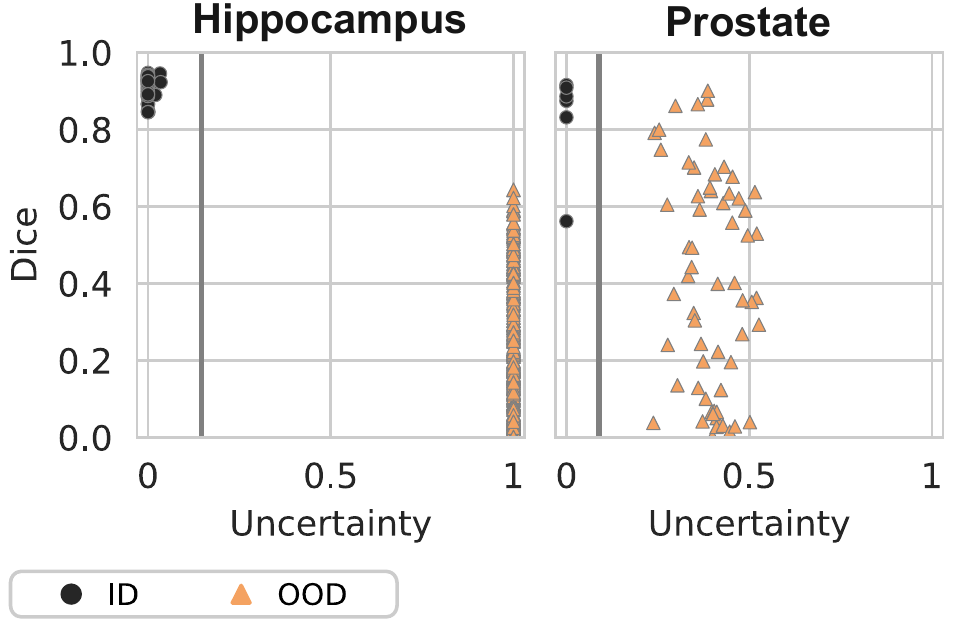}
\caption{Dice coefficient against normalised uncertainty for the segmentation of the hippocampus (left) and prostate (right) in MR images. Black circles are test ID (\emph{MSD}) images, and orange triangles are OOD cases.} \label{fig:mrtscatter}
\end{figure}

Table \ref{tab:hip_prost} summarises our results on OOD detection, and we visualise the uncertainties of our method in Figure \ref{fig:mrtscatter}. We immediately see that -- for both MR segmentation tasks -- detecting OOD cases is much easier than for chest CT. In all cases, the proposed method correctly distinguishes ID from OOD data. This is likely due to the inherent variability across MRI datasets in terms of intensity histogram and fields-of-view. The last row includes a \emph{far-OOD} case where we look to detect \emph{MSD H} cases on the model trained with \emph{MSD P} and vice versa. This also seems to be an easy problem, and our method correctly identifies all OOD cases.

\begin{table}[ht]
\centering
\begin{adjustbox}{max width=\linewidth}
{\begin{tabular}{lllll}
\toprule
\textbf{\textbf{Method}} & \textbf{ESCE $\downarrow$} & \textbf{Error $\downarrow$} & \textbf{FPR $\downarrow$} & \textbf{AUC $\uparrow$}\\
\midrule
Max. Softmax & .20/.36 & .05/.49 & \textbf{.00}/.82 & \textbf{1.0}/.74\\
MC Dropout $N=10$ & .53/\textbf{.08} & .50/.01 & 1.0/.02 & .40/\textbf{1.0}\\
MC Dropout $N=100$ & .48/.14 & .53/\textbf{.00} & 1.0/\textbf{.00} & .12/\textbf{1.0}\\
KL & .18/.15 & .05/.16 & \textbf{.00}/.16 & \textbf{1.0}/.83\\
TTA & .20/.40 & .09/.25 & \textbf{.00}/\textbf{0.0} & \textbf{1.0}/.83\\
Temp. Scaling & \textbf{.12}/.36 & .03/.49 & \textbf{.00}/.82 & \textbf{1.0}/.74\\
Energy Scoring & .68/.53 & .50/.49 & 1.0/.98 & .50/.12\\
\textbf{Ours} & .21/.19 & \textbf{.00}/\textbf{.00} & \textbf{.00}/\textbf{.00} & \textbf{1.0}/\textbf{1.0}\\
\midrule
Ours far-OOD & .08/.01 & .00/.00 & .00/.00 & 1.0/1.0\\
\bottomrule
\end{tabular}}
\end{adjustbox}
\caption{MRI results. Segmentation calibration (as ESCE) and OOD detection scores between test ID and OOD cases for hippocampus and prostate, respectively. The networks were trained with \emph{MSD H} and \emph{MSD P} data, respectively, so these cases are ID. The last row summarises the results for the far-OOD case of detecting \emph{MSD P} cases on the \emph{MSD H} model and vice versa.}
\label{tab:hip_prost}
\end{table}

\section{Discussion}

Uncertainty quantification is an unavoidable cornerstone for safely deploying predictive models in real clinics. Our results show that the proposed distance-based approach provides valuable information for detecting images that the model is unprepared to segment.

As distance-based OOD detection can seamlessly augment any segmentation pipeline, there is no reason against performing this quality check. However, we found in our analysis several areas where there is room for improvement. Almost all our experiments showed that our method is overly cautious in its uncertainty estimation. Specifically, many OOD cases for which the model \emph{did} produce adequate segmentation were deemed highly uncertain. Only for the \emph{artefact shift} scenario were weekly transformed samples segmented.

The \emph{artefact} and \emph{affine shifts} experiments show that -- for both explored synthetic scenarios -- the produced distances grow linearly with the degree of change and are inversely proportional to segmentation quality. This is ideal behaviour for an uncertainty metric. However, the same does not hold for the \emph{dataset shift} and \emph{diagnostic shift} settings. Particularly for the last scenario, our method assigns similar uncertainties to both Covid-19 and non-Covid cases, even though segmentations are much worse for the last group. Further research should explore which distribution shifts negatively affect model performance, and how these can be distinguished from harmless shifts.

This discrepancy might also be associated with the relatively higher variety of the pulmonary patterns for the labels GGO and consolidation present in the various pulmonary diseases making up the non-Covid-19 group, as compared to the Covid-19 group. This group was, however, purposefully designed to resemble a broad range of non-Covid-associated pulmonary disease patterns, which represent Covid-19-mimics. Further, the large time frame in which these cases were collected, as well as a differing distribution amongst the three CT scanners used to generate these cases, might contribute to this finding.

Our experiments also show that our distance-based approach does not adequately detect poorly segmented cases for in-distribution data. This shortcoming reinforces the notion that uncertainty estimation methods, which are mainly designed to detect uncertain predictions in ID data, should complement OOD detection in practice. However, neither MC Dropout nor TTA were successful at assessing segmentation quality.

Our ablation study shows that intermediate network layers are the most informative for assessing distribution shifts. OOD samples do not display patterns that differ sufficiently from training samples in feature maps near the inputs or outputs of the model. In contrast, activations in intermediate layers allow the separation between ID and OOD cases. For the HighResNet model, which does not follow an encoder-decoder structure, dilated convolutions near the end of the model resulted in the best uncertainty estimates.

Finally, our \textit{far-OOD} experiments on both CT and MR data confirm that our proposed method accurately detects cases very far from the training distribution. Such \textit{far-OOD} cases may arise when an erroneous input is fed into the model, and automatically signalling such mistakes can be helpful for inexperienced users.
\section{Conclusions}

Despite ample progress in the development of segmentation solutions, these are not ready to be deployed in clinical practice. The main reason behind this is the fact that predictive models fail silently, coupled with a lack of appropriate quality controls to detect such behaviour. This is particularly true when it is not trivial to identify a faulty output, such as segmentation of SARS-CoV-2 lung lesions.

Increasingly, institutions are taking part in initiatives to gather large amounts of annotated, heterogeneous data and release it to the public. This could allow the training of robust models and potentially alleviate the burden of radiologists. However, even models trained with heterogeneous cohorts are susceptible to distribution shifts.

We propose a distance-based method to detect images far from the training distribution in a low-dimensional feature space, and find that this is a lightweight and flexible way to signal when a model prediction should not be trusted.

Future work should explore how to improve uncertainty calibration by identifying high-quality predictions. For now, our work increases clinicians' trust while translating trained neural networks from challenge participation to real clinics.

\section*{Acknowledgments}
This work was supported by the RACOON network under BMBF grant number [01KX2021];
and the Bundesministerium f\"{u}r Gesundheit (BMG) with grant [ZMVI1-
2520DAT03A].

\bibliographystyle{model2-names.bst}\biboptions{authoryear}
\bibliography{refs}

\end{document}